\documentclass[numberedappendix,apj]{emulateapj}

\usepackage{amsmath}
\slugcomment{To be submitted}
\shortauthors{Bernet et. al.}

\begin{document}
\title{MgII absorption systems with $W_{0} \ge 0.1 \rm{\AA}$ 
for a radio selected sample of 77 QSOs and their associated 
magnetic fields at high redshift\footnote{Based on observations made with ESO telescopes at the
    Paranal Observatory under Programme IDs 075.A-0841 and 076.A-0860}}

\author{M. L. Bernet,  F. Miniati, S.J. Lilly}
\affil{Physics Department, ETH Zurich, Wolfgang-Pauli-Strasse 27, CH-8093 Zurich, Switzerland}
\email{mbernet@phys.ethz.ch, fm@phys.ethz.ch, simon.lilly@phys.ethz.ch}

\begin{abstract}
  We present a catalogue of MgII absorption systems obtained from high
  resolution UVES/VLT data of 77 QSOs in the redshift range $0.6 < z <
  2.0$, and down to an equivalent width $W_{0} \ge 0.1
  \rm{\AA}$. The statistical properties of our sample are found to be
  in agreement with those from previous work in the
  literature. However, we point out that the previously observed
  increase with redshift of $\partial N / \partial z$ for weak
  absorbers, pertains exclusively to very weak absorbers with $W_{0}
  <0.1 \rm{\AA}$. Instead, $\partial N / \partial z$ for absorbers
  with $W_{0}$ in the range 0.1--0.3 $\rm{\AA}$ actually decreases
  with redshift, similarly to the case of strong absorbers. 
  We then use this catalogue to extend our earlier analysis of the
  links between the Faraday Rotation Measure of the quasars and the
  presence of intervening MgII absorbing systems in their spectra. In
  contrast to the case with strong MgII absorption systems ($W_{0} >
  0.3 \rm{\AA}$), the weaker systems do not contribute significantly
  to the observed Rotation Measure of the background quasars. This is
  possibly due to the higher impact parameters of the weak systems
  compared to strong ones, suggesting that the high column density
  magnetized material that is responsible for the Faraday Rotation is
  located within about 50 kpc of the galaxies.
  Finally, we show that this result also rules out the possibility
  that some unexpected secondary correlation between the quasar
  redshift and its intrinsic Rotation Measure is responsible for the
  association of high Rotation Measure and strong intervening MgII
  absorption that we have presented elsewhere, since this would have
  produced an equal effect for the weak absorption line systems, which
  exhibit a very similar distribution of quasar redshifts.
\end{abstract}


\section{Introduction} \label{introduction:sec}

Quasar absorption line systems provide a unique tool to study the
evolution of galaxies through their lifetime. Since the galaxies
associated with such systems are selected by their gas cross section,
as compared to their stellar luminosities from multi broad band imaging 
techniques, absorption lines provide us with a
complementary view on galaxy properties and evolution. One of the
best studied population of galaxies is the one selected by the MgII
absorption doublet $\lambda\lambda 2796.35,2803.53$ \AA.  
Magnesium is produced by alpha-processes in post main sequence stars
and, therefore, is abundant in galaxies. In addition, MgII doublet
is easy to observe because it has a large cross section for absorption,
it can be easily identified and the rest frame wavelengths $\lambda\lambda
2796.35,2803.53$~\AA~ are detectable from the ground over the full
redshift range $0.3 < z < 2.2$. 

Depending on whether the equivalent width, $W_{0}$ of the MgII
2796~\AA~line is greater or smaller than $0.3 \rm{\AA}$, the
absorption systems is classified as strong or weak, respectively.  The
fact that the equivalent width distribution appears to steepen below
$0.3 \rm{\AA}$, lends support to the idea that weak and strong
absorbers constitute two distinct populations.  (Nestor et al. 2005,
Narayanan et al. 2007).

It is well established that if there is strong MgII absorption line,
then in almost all cases a galaxy is found within 100 kpc with a median
impact parameter of around 50 kpc (Steidel et al. 1995, 
Churchill et al. 2005, Zibetti et al. 2007, Kacprzak et al. 2008, 
Chen $\&$ Tinker 2008). 

The MgII host galaxies are known to span a broad range of optical
luminosities ($0.1 -3.0 L^{\star}$) and colors (Bergeron et al. 1986,
Steidel 1995, Kacprzak et al. 2008).  The gas traced by strong
MgII absorption spans a wide range of neutral hydrogen column
densities from $10^{17}$ to $10^{22} \rm{cm^{-2}}$ (Churchill et
al. 2000, Rao et al. 2006). Recently M{\'e}nard \& Chelouche 2008
found a strong correlation between the median $N_{HI}$ and the MgII
rest equivalent width $W_{0}$.

However, the nature of the objects selected by weak MgII absorption is
not yet completely clear.  It has been suggested that they form a
separate populations of galaxies with respect to those traced by
strong absorbers, e.g. low surface brightness galaxies or dwarf
galaxies (Churchill et al. 1999, Narayanan et al. 2007).
Alternatively, the weak systems could be tracing the same galaxies as
the strong absorbers, but at a higher QSO impact parameter where the
gas density is much lower and, hence, the equivalent width smaller
(Churchill et al. 2005, Kacprzak et al. 2008).

In this paper we present observations obtained with the Ultraviolet
and Visual Echelle Spectrograph (UVES)(Dekker et al. 2000) at the VLT
and the associated catalogue of strong and weak MgII absorption lines
down to an equivalent width limit of $0.1 \rm{\AA}$, giving details of
the selection criteria and the methods applied. We also construct the
inferred number densities of MgII system and the observed equivalent
width distribution and compare these with results published in the
literature. Compared with most previous surveys for strong MgII
absorption systems (Lanzetta et al. 1987, Steidel et al. 1992, Nestor
et al. 2005, Nestor et al. 2006) we have a much higher spectral
resolution, $R\approx 43000$ which allows us to identify the strong
MgII systems unambigously.

In a previous paper we have used the catalogue of strong MgII
absorbers presented in this paper to probe the magnetic fields in
normal galaxies at redshift $z\sim 1$ (Bernet et al. 2008). As we had
already hypothetized in Kronberg et al. (2008) in order to eplain the
observed increase in QSO RM as a function of the QSO redshift,
in Bernet et al. (2008) we demonstrated that lines of sight with
strong MgII absorption lines have significantly higher RM than
those without. This implies the presence of $\mu$G strong, 
large scale magnetic fields in the associated galaxies. 

In this work we extend our analysis to weak MgII absorption systems
with equivalent width in the range $0.1\rm{\AA} -0.3 \rm{\AA}$.  We
show that unlike their strong counterpart, weak absorbers do not
contribute with any significance to the observed RMs. We discuss the
possible interpretations of this result. We also show that this result 
argues against the case in which the correlation between RM and strong
MgII absorption systems reported in Bernet et al. (2008) arises 
due to an intrinsic evolution of QSOs magnetic fields.

The paper is organized as follows: in Sec.~\ref{data:sec} we present
the MgII absorption system catalogue, including the details of the
observations and data analysis; in Sec.~\ref{mgII:sec} we analyze the
statistical properties of the detected absorption systems and compare
with previous work; Sec.~\ref{RM:sec} we investigate the correlation
of the MgII system with RM; a short summary in Sec. \ref{Sum:sec}
concludes the paper.

\section{Optical Data} \label{data:sec}

\subsection{Spectra}

Our dataset consists of 77 QSO spectra obtained with the high
resolution UVES spectrograph at the VLT. The QSOs were selected from
the larger sample of 901 radio sources with determined RM and
redshifts presented in Kronberg et. al (2008). The selection was based
on the following criteria: i) redshift range $0.6 < z_{QSO} < 2.0$,
ii) Galactic latitudes $\left|b\right| > 30^{\circ}$ and iii) $m_{V} <
19$. The latter was imposed to have adequate S/N in exposures times up
to 30 minutes per object.  In order to produce a complete census of
strong MgII absorption systems in the redshift range $0.35 < z <
z_{QSO}$, we required a signal to noise ratio per resolution element
S/N$\ge 10$, across the full spectral range, although the data quality
was typically better than that.  To fully cover the redshift path to
each QSO we chose the following standard setting of UVES, with
dichroic 1 having central wavelengths 390nm and 580nm, and dichroic 2
having central wavelengths 437nm and 760nm for the blue and red arm,
respectively. This results in complete wavelength coverage from
3480$\rm{\AA}$ to 9460 $\rm{\AA}$ with only two small spectral gaps
between 5762-5834 $\rm{\AA}$ and 7513-7660 $\rm{\AA}$ due to the small
gap between the two CCDs of the red arm of UVES.  We chose a slit
width of $1''$ corresponding to $R\approx45000$ and $R\approx43000$ in
the blue arm and red arm respectively.  The observations took place
during 3 nights on 28-30 July 2006 (in visitor mode) and during 30 hrs
of service mode observations carried out between October 2006 to April
2007.

\subsection{Data reduction}

The spectra were bias subtracted, flat-fielded and wavelength
calibrated using the ESO MIDAS package in the OPTIMAL mode. The reduced 1
dimensional vacuum-heliocentric corrected exposures were first scaled
to the same flux level and then co-added, weighted by the S/N of the
corresponding pixel. The spectra were then normalized using a spline
method to fit the continuum. To increase the S/N of the spectra the
overlapping wavelength regions of different settings were also
co-added.

To detect the absorption lines we used the so called ``aperture''
method (Lanzetta et al. 1987). Namely, we identified strong MgII
absorption systems by requiring a 5$\sigma$ detection of the rest
frame equivalent width of the MgII $\lambda$ 2796 \AA~ line. The
absorption system was considered a MgII absorption identification if
there was also at least a 2.5$\sigma$ detection of the corresponding
$\lambda$ 2803 \AA~ line at the same redshift. We further checked if
the doublet ratio, $DR=\frac{EW_{2796}}{EW_{2803}}$, of the MgII lines
lay within the range from 1.0 for completely unsaturated lines and 2.0
for saturated lines (within errors). We also visually checked the line
system for similar profile shapes to exclude chance alignments. MgII
systems within 500 km $s^{-1}$ of each other are considered as one
single absorption system with the equivalent widths added together. We
also visually inspected the flux spectra to look for MgII absorption
lines that might have been missed by the alogrithm. This was twice the
case, where the stronger $\lambda 2796$ \AA~ line was detected as a
single absorption line but the weaker $\lambda 2803$ \AA~ line
consisted of 2 separate lines, with miscalculated centroids.

\begin{table*}
    \centering
   % \begin{minipage}{\textwidth}
    \caption{QSOs with strong MgII absorption systems}
        \begin{tabular}{llllllll}%{l|l|l|l}
        \hline
        \hline
        & & & & & & &\\
        QSO &$z_{QSO}$ &  RA & Dec &$z_{MgII}$ & $W_{0}(2796)$ & $z_{MgII}$ & $W_{0}(2796)$ \\

       & & (J2000) & (J2000) & & (\AA) & & (\AA) \\
        (1) & (2) &(3) & (4) & (5)& (6) &(7)& (8)\\
        & & & & & & & \\
        \hline
      & & & & & & & \\
        PKS1244-255 & 0.638 & 12:46:46.8 & -25:47:49 &  0.49286 & 0.68 & & \\
 % & & & & &\\
OX-192 & 0.672 &21:58:06.3 &-15:01:09 & 0.63205 & 1.40 & & \\
% & & & & &\\
4C+19.44 & 0.72 &13:57:04.4 &+19:19:07 & 0.45653 & 0.85 & & \\
% & & & & &\\
OC-65 &0.733 & 01:41:25.8 & -09:28:44 &0.50046 & 0.53 & & \\
% & & & & &\\
4C+19.34 & 0.828 & 10:24:44.8 &+19:12:20 &0.52766 & 1.00 & & \\
% & & & & &\\
PKS0420-01 & 0.915 & 04:23:15.8 &-01:20:33 & 0.63291 & 0.77 & & \\
3C336$^{a}$ & 0.9274 &16:24:39.1 &+23:45:12 & 0.47192 & 0.93 &0.36811  &0.19    \\
& & & &0.65607 & 1.45 & 0.51742 & 0.13 \\ 
 & & & &0.79707 & 0.46 & & \\
 & & & &0.89119 & 1.58 & & \\
% & & & & &\\
PKS2354-11$^{b}$& 0.96 &23:57:31.2 & -11:25:39 & 0.54456 & 0.53 &0.3715 & 0.10\\
                & & & & & &0.40704 &0.12  \\
               & & & & & & 0.56496 & 0.24 \\
PKSB1419-272 & 0.985 &14:22:49.2 &-27:27:56 & 0.55821 & 0.44 & 0.4319 & 0.26 \\
% & & & & &\\
4C+6.69 & 0.99 &21:48:05.4 & +06:57:39 & 0.79086 & 0.55 & & \\
% & & & & &\\
4C+01.24 & 1.018 & 09:09:10.1  &+01:21:36  & 0.53587 & 0.44 & & \\
% & & & & &\\
PKS0130-17 & 1.022 & 01:32:43.5 &-16:54:49 &0.50817 & 0.59 &0.82200 &0.28 \\
   & & & & & &0.86895 &0.28 \\
% & & & & &\\
4C-02.55 & 1.045 &12:32:00.0 &-02:24:05 & 0.39524 & 2.03 & 0.83083 & 0.12 \\
      & & & &0.75689 & 0.30 & & \\
% & & & & &\\
MRC0122-003 & 1.07 & 01:25:28.8 & -00:05:56 & 0.39943 & 0.47 &0.9534 & 0.2 \\
% & & & & &\\
PKS0506-61 & 1.093 &05:06:43.9 &-61:09:41 & 0.92269 & 0.49 & & \\
% & & & & &\\
PKS0426-380$^{c}$ & 1.11 &04:28:40.4 &-37:56:20 & 0.55855 & 0.93 & & \\
   & & & &1.02886 & 0.57 & & \\
3C208 & 1.11 & 08:53:08.8 &+13:52:55 & 0.65262 & 0.62 & & \\
   & & & &0.93537 & 0.40 & & \\
% & & & & &\\
4C+13.46 & 1.141 &12:13:32.1 &+13:07:21 & 0.77189 & 1.29 & 0.37536 & 0.11 \\
% & & & & &\\
PKS0038-020 & 1.178 & 00:40:57.6 & -01:46:32 & 0.68271 & 0.35 &0.40701 &0.21 \\
  & & & & & &1.00936 & 0.14\\
% & & & & &\\
PKS2204-54 & 1.206 &22:07:43.7 &-53:46:34 & 0.6877 & 0.73 &0.43723 &0.29 \\
    & & & & & &0.97152 & 0.15 \\
% & & & & &\\
4C+06.41 & 1.27 &10:41:17.1 & +06:10:17 &0.44151 & 0.69 &0.65536 &0.07 \\
% & & & & &\\
PKS0839+18 & 1.27 & 08:42:05.1 &+18:35:41 & 0.71118 & 0.56 &0.63087 & 0.10 \\
% & & & & &\\
PKS2326-477 & 1.299 & 23:29:17.7 &-47:30:19 & 0.43195 & 0.38 & & \\
   & & & &1.26074 & 0.66 & & \\
% & & & & &\\
PKS1615+029 & 1.339 &16:17:49.9 &+02:46:43 & 0.52827 & 0.31 & & \\
% & & & & &\\
PKS0112-017 & 1.365 & 01:15:17.1 &-01:27:05 & 1.18965 & 0.90&0.95289 &0.14 \\
% & & & & &\\
PKS2223-05 & 1.404 & 22:25:47.2 &-04:57:01 &0.84652 & 0.60 & & \\
% & & & & &\\
PKS0402-362 & 1.417 & 04:03:53.7 &-36:05:02 & 0.79688 & 1.80 & & \\
% & & & & &\\
PKS0332-403 & 1.445 & 03:34:13.7 &-40:08:25 & 1.20898 & 0.79 & & \\
% & & & & &\\
OQ+135 & 1.611 &14:23:30.1 & +11:59:51 & 1.36063 & 0.51&0.61474 &0.12 \\
% & & & & &\\
% & & \\
OX+57 & 1.932 & 21:36:38.6 & +00:41:54 & 0.62855 & 0.60 & 0.81024 &0.1 \\
% & & & & &\\
OW-174 & 1.932 &20:47:19.7 &-16:39:06 &1.32871 & 0.61 & 0.83407 &0.22 \\
    & & & & & &1.34258 &0.07 \\
% & & & & &\\
PKS1143-245 & 1.95 & 11:46:08.1 & -24:47:33& 1.24514 & 0.30 &0.41966 & 0.16 \\
   & & & &1.52066 &0.46 & 0.87633&0.07 \\
4C+5.81$^{a}$& 1.967 &21:53:24.7 &+05:36:19 & 1.88286 & 0.89 & & \\   
% & & & & &\\
PKS1157+014 & 1.986 &11:59:44.8 & +01:12:07 & 1.94372 & 1.58 &0.79082 & 0.09 \\
   & & & & & &1.3305  & 0.11\\
% & & & & &\\
PKS2353-68 & 2.77 & 23:56:00.7 & -68:20:03 & 1.26958 & 0.40 &1.50565 & 0.13 \\
  & & & &1.85808 &0.58 & & \\
        \hline 
        \end{tabular}
    \label{tab:strongMgII}
  %  \end{minipage}
\begin{flushleft}
COLUMNS. - (1) Name of the source, (2) Redshift of the source, (3),(4) Coordinates, (5) Redshifts of the systems with strong MgII absorption, (6) Equivalent widths $W_{0}(2796)$ of the strong MgII absorption systems, (7) Redshifts of the systems with weak MgII absorption, (8) Equivalent widths $W_{0}(2796)$ of the weak MgII absorption systems  
\\
 COMMENTS. -  $^{a}$Sources were not included in the work of Bernet et al. 2008 because the optical and the radio emission are seperated by more than 5''. $^{b}$Source was not included in the work of Bernet et al. 2008 due to MgII absorption local to the QSO at $z_{MgII}=0.9587$ with $W_{0}=0.61$. $^{c}$ This source was not used in Bernet et al. 2008 due to a misidentification.\\

\end{flushleft}
\end{table*}

\begin{table*}
    \centering
    \caption{QSOs with MgII absorption systems in the range $0.1 \rm{\AA} \le W_{0}(2796) < 0.3 \rm{\AA}$  }
    \begin{tabular}{llllll}%{l|l|l|l}
        \hline
        \hline
        & & & \\
        QSO & $z_{QSO}$ & RA & Dec & $z_{MgII}$ & $W_{0}(2796)$\\
        & &(J2000) &(J2000) & & (\AA) \\
        (1) & (2) & (3) & (4) & (5) & (6)\\
        & & & \\
        \hline

 & & & & & \\
OC-192$^{a}$ & 0.616 &01:57:41.6 &-10:43:40 &0.42036 &0.11\\
% & & \\
3C095 & 0.616 &03:51:28.5 &-14:29:09 &0.35717 &0.12\\
% & & \\
3C037 & 0.672 & 01:18:18.5 &+02:58:06 &0.39178 &0.17 \\
% & & \\
OB-94 & 0.719 & 00:59:05.5 &+00:06:52&0.42816 &0.08 \\
 & & & &0.48866 &0.11 \\
% & & \\
PKS1424-11 & 0.805 &14:27:38.1 &-12:03:50 &0.65781 &0.25 \\
% & & \\
OC-259 & 0.837 &  01:37:38.3 &-24:30:54 &0.46592 &0.15 \\
% & & \\
PKS1111+149 & 0.869 &11:13:58.7 &+14:42:27 &0.64557 &0.16 \\
% & & \\
PKS2340-036 & 0.896 & 23:42:56.5 &-03:22:26 &0.421 &0.16 \\
 & & & &0.68381 &0.17 \\
% & & \\
PKS2255-282$^{a}$ & 0.926 &22:58:05.9 &-27:58:21 &0.60402 &0.26 \\
    & & & &0.84972 &0.19 \\
% & & \\
OM+133$^{b}$  & 1.040 &11:22:29.7 &+18:05:26 &1.00946 &0.17 \\
% & & \\
3C245 & 1.029 &10:42:44.6 &+12:03:31 &0.65969 &0.16 \\
% & & \\
4C11.69 & 1.037 &22:32:36.4 &+11:43:51 &0.73783 & 0.18 \\
% & & \\
4C20.24 & 1.11 &10:58:17.9 &+19:51:51 &0.85629 &0.13 \\
% & & \\
4C-5.64 & 1.185 &15:10:53.6  &-05:43:07 &0.87758 &0.13 \\
% & & \\
PKS2335-18 & 1.446 &23:37:56.6 &-17:52:20 &0.57337 &0.27 \\
% & & \\
PKS0202-17 & 1.74 &  02:04:57.7&-17:01:20 &0.51833 &0.2 \\
% & & \\
4C-4.04$^{c}$ & 1.925 &01:22:27.9 &-04:21:28 &0.65744 & 0.29\\
  & & & &0.71871 &0.16 \\
  & & & &0.91435 &0.23 \\
\hline

\end{tabular}

    \label{tab:weakMgII}
\begin{flushleft}
COLUMNS. - (1) Name of the source, (2) Redshift of the source, (3),(4) Coordinates, (5) Redshifts of the systems with strong MgII absorption, (4) Equivalent widths $W_{0}(2796)$ of the strong MgII absorption systems, (6) Redshifts of the systems with weak MgII absorption, (7) Equivalent widths $W_{0}(2796)$ of the weak MgII absorption systems  
\\
COMMENTS. - $^{a}$Sources show weak MgII absorption lines local to the QSOs. $^{b}$Source was not included in the work of Bernet et al. 2008 due to MgII absorption local to the QSO at $z_{MgII}=1.0439$,$W_{0}=1.27$. $^{c}$Spectra shows no strong intervening MgII absorption lines and was included in Bernet et al. 2008 but a recheck showed a broad MgII absorption system at $z=1.96429$ with $W_{0}=0.39$ and should be therefore excluded from this work.\\

\end{flushleft}
\end{table*}

\subsection{Completeness} \label{compl:sec}
\begin{figure}[tbh] 
\includegraphics[width=1.0\hsize]{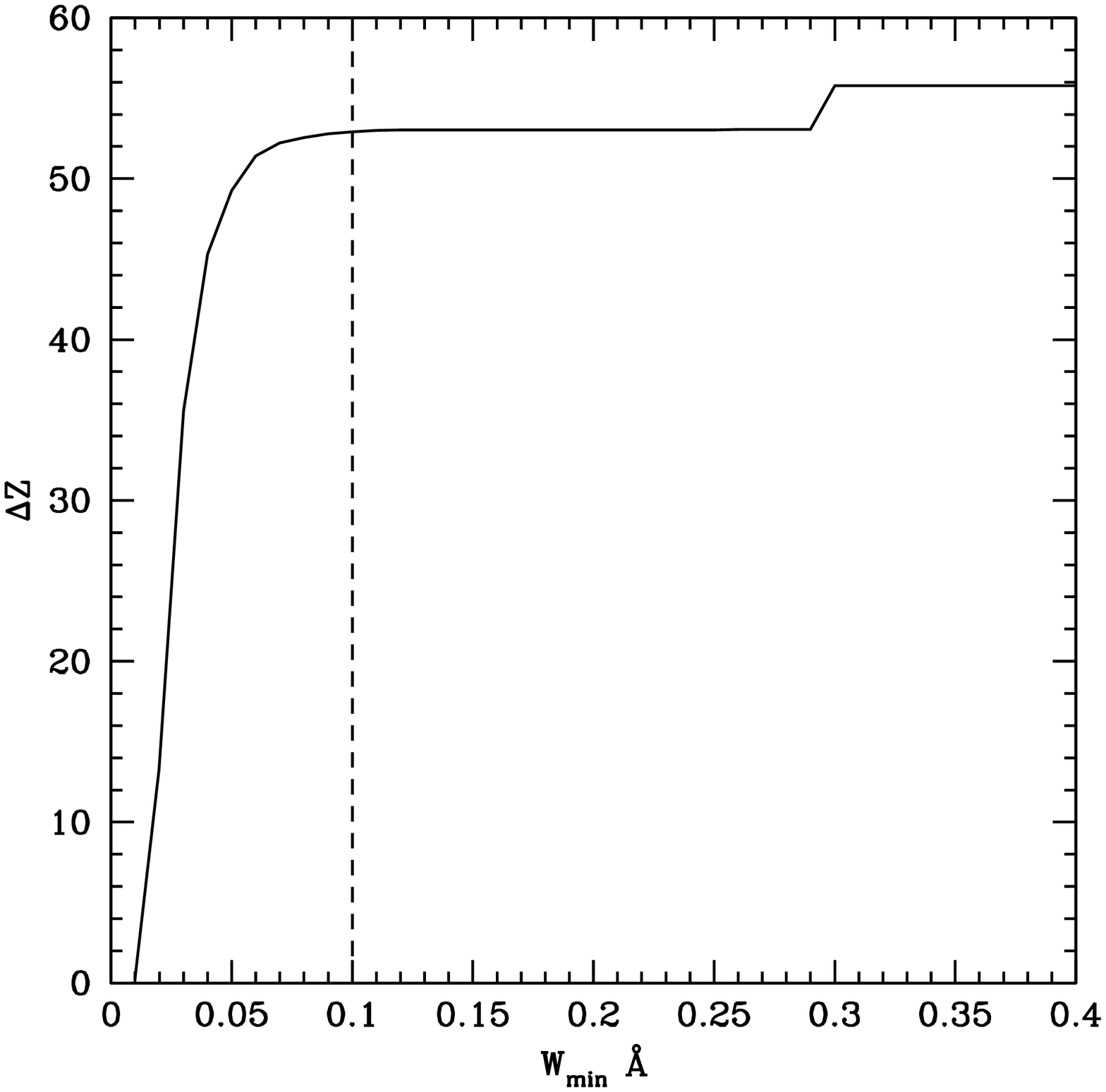}
\caption{Covered redshift path $\Delta Z$ as a function of the
  rest-frame equivalent width detection threshold of the 2796 line,
  $W_{min}$, for a $5\sigma$ significance level.  \label{fig:zpath}}
\end{figure}

To check the completeness of our catalogue, following
Lanzetta et al. (1987) we calculated the redshift path of
the survey for a certain equivalent width detection threshold
$W_{min}$.  The redshift path is given by
\begin{equation}
\Delta Z(W_{min})=\int_{z_{min}}^{z_{max}}\sum_{i}^{N_{los}}g_{i}(W_{min},z)dz, 
\end{equation}
where
\begin{center}
\begin{eqnarray}
 g_{i}(W_{min},z)=&H&(z-z_{i}^{min})H(z_{i}^{max}-z)\times  \\
\nonumber
 &H&(W_{min}-N_{\sigma}\sigma_{EW}(z)/(1+z))
 \label{eq:Heavyside}
 \end{eqnarray}

\end{center}

is 1 if the MgII $\lambda \lambda$ 2796,2803 doublet could have been
detected at redshift z with a rest frame equivalent width greater or
equal to $W_{min}(2796)$ at a significance level greater than
$N_{\sigma}$, and zero otherwise. In eq. 2 % \ref{eq:Heavyside}
, $H()$
is the Heavyside function, $z_{i}^{min}$ and $z_{i}^{max}$ are the
minimum and maximum redshift observed for the i-th QSO, that is 0.345
and the quasar redshift, $z_{QSO}$, respectively. Also, $\sigma_{EW}(z)$
is the Poisson error on the equivalent width per resolution element at
redshift z. 
\begin{table*}
    \centering
    \caption{QSOs with no MgII absorption systems or below  $W_{0}(2796) < 0.1 \rm{\AA}$}
   \begin{tabular}{llllll}%{l|l|l|l}
        \hline
        \hline
        & & & & &\\
        QSO & $z_{QSO}$ & RA & Dec &$z_{MgII}$ & $W_{0}(2796)$  \\
        & &(J2000) &(J2000) & & (\AA) \\
        (1) & (2) & (3) & (4) & (5) & (6)\\
        & & & & &\\
        \hline
        & & & & &\\
3C281 & 0.602 &13:07:54.0 &+06:42:14 & &\\
% & & \\
4C-06.35 & 0.625 & 13:38:08.1 & -06:27:11 & &\\
% & & \\
PKS2243-123 & 0.63 &22:46:18.2 &-12:06:51 & &\\
% & & \\
4C+02.27 & 0.659 &09:35:18.2 &+02:04:16 & &\\
% & & \\
3C057 & 0.669 &  02:01:57.1&-11:32:33 & &\\
% & & \\
OX-173 & 0.701 &21:46:23.0 &-15:25:44 & &\\
% & & \\
3C039 & 0.765 & 01:21:01.2  & +03:44:14 & &\\
% & & \\
PKS0414-06 & 0.775 &04:17:16.7 &-05:53:45 & &\\
% & & \\
PKS0422-380 & 0.782 & 04:24:42.2 &-37:56:21 & &\\
% & & \\
3C454.3 & 0.859 &22:53:57.7 & +16:08:54& & \\
% & & \\
ON+187 & 0.871 &12:54:38.2 &+11:41:06 & &\\
% & & \\
4C-00.50 & 0.89 &13:19:38.7 & -00:49:40 & &\\
% & & \\
OD+094.7 & 0.893 & 02:59:27.1 & +07:47:40  & &\\
4C+14.31$^{a}$& 0.896 &09:25:07.3 &+14:44:26 & & \\ 
% & & \\
4C-3.79 & 0.901 &22:18:52.0 &-03:35:37 & &\\
% & & \\
TXS0223+113 & 0.924 & 02:25:41.9  &+11:34:25 & &\\
% & & \\
3C094 & 0.962 & 03:52:30.5 &-07:11:02 & &\\
% & & \\
OX-325 & 0.979 &21:18:10.6 &-30:19:12 &0.4271 &0.04 \\
% & & \\
PKS1127-14 & 1.187 &11:30:07.1 &-14:49:27 & &\\
% & & \\
4C-5.62$^{b}$ & 1.249 &14:56:41.4 &-06:17:43 & &\\
% & & \\
4C+5.64 & 1.422 &15:50:35.3 & +05:27:10 & &\\
% & & \\
3C298$^{b}$ & 1.436 & 14:19:08.2&+06:28:35 &1.27336 &0.06 \\
% & & \\
OD-055 & 1.45 &02:35:07.4 &-04:02:06 & &\\
% & & \\
OK186 & 1.472 & 09:54:56.8 & +17:43:31& &\\
% & & \\
PKS2227-08 & 1.559 &22:29:40.1 &-08:32:54 &0.63894 &0.07 \\
\hline

 		\end{tabular}

    \label{tab:noMgII}
\begin{flushleft}
COLUMNS. - (1) Name of the source, (2) Redshift of the source, (3) Redshifts of the systems with weak MgII absorption, (4) Equivalent widths $W_{0}(2796)$ of the weak MgII absorption systems  
\\
 COMMENTS. - $^{a}$Source was not included in the work of Bernet et al. 2008 because the optical and the radio emission are seperated by more than 5''. Source shows MgII absorption local to the QSO at $z=0.88123$ with $W_{0}=1.15$. $^{b}$Sources show weak MgII absorption lines local to the QSOs.\\
\end{flushleft}
\end{table*}

The redshift path of the survey $\Delta Z$ as a function
of the equivalent width limit $W_{min}$ is shown in Figure
$\ref{fig:zpath}$. It can be seen that the redshift path begins to
fall below 0.1 $\rm{\AA}$ but is nearly constant above, showing
that our sample has a high completness down to $W_{min}\ge 0.1$~\AA.
Finally, the drop in $\Delta Z$ at $W_{min}=0.3 \rm{\AA}$ is because
for the weak MgII systems we excluded spectral ranges known for
atmospheric absorption from the search.

We miss only about $\Delta Z=1.5$ of the total available redshift range between $z=0.345$ and the quasar redshifts $z_{QSO}$ 
(due to the two small spectral gaps). This leads to a high completness of the redshift coverage of $97\%$ at $W_{min}=0.3 \rm{\AA}$ and $92\%$ at $W_{min}=0.1 \rm{\AA}$.

\subsection{MgII Absorbers Catalogues}
Our sample includes a total of 44 strong MgII absorption systems with
mean redshift $\langle z_{MgII} \rangle=0.85$ and a total redshift
path of $\Delta Z=55.8$, and 44 weak MgII systems in the
equivalent width range 0.1 - 0.3$\rm{\AA}$, with a redshift path of
$\Delta Z=53.0$.  The complete catalogue of detected strong MgII
absorption systems is presented in Table \ref{tab:strongMgII}.
However, MgII absorption lines lying within 3000 km/s of the quasar
redshift, i.e. arising in the environment local to the QSO, were
excluded in any further analysis to maintain the sample of MgII host
galaxies as homogenous as possible.  When multiple systems of
absorbers are detected along the line of sight to the same QSO, the
equivalent width and redshift for each one of them is provided. When
weak absorbers are present, they are also listed but in a separate
column, with the caveat of both incompleteness and non-systematic
search below for $W_{0} < 0.1$~\AA.  Finally, a catalogue of weak
absorbers with $W_{0} \ge 0.1$~ \AA~ is presented in
Table~\ref{tab:weakMgII}, and the remaining observed lines of sight
are reported in Table~\ref{tab:noMgII}.

\subsection{Note on the QSO PKS 2353-68}
\begin{figure}
\includegraphics[angle=270,width=1.0\hsize]{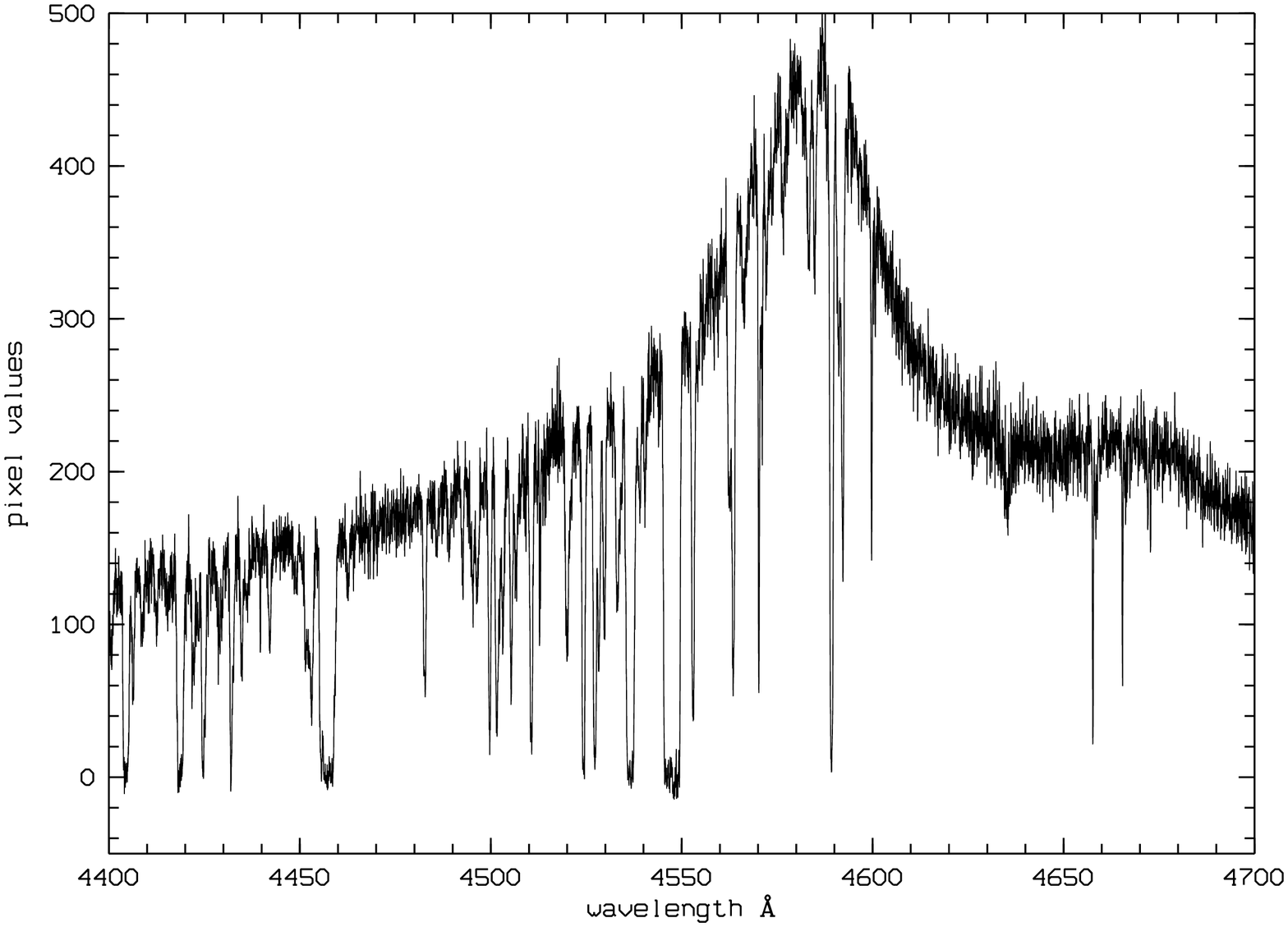}
\caption{Broad Lyman-$\alpha$ emission of PKS 2353-68 at z=2.77. At
  wavelength bluewards of the Lyman-$\alpha$ emission the
  Lyman-$\alpha$ forest can be seen. This object was given before a
  redshift of $z=1.716$ (Peterson \& Bolton 1972), but actually the
  object with $z=1.716$ lies 115'' away from the radio source PKS
  2353-68.\label{fig:Lyalpha}}
\end{figure}

The high redshift QSO PKS 2353-68 is part of our sample and the
optical spectra allowed us to make an independent estimate of its
redshift, which disagrees with the value of $z=1.716$ from the Hewitt
\& Burbidge (1989) optical catalague. In fact, the radio emission of
this QSO is centered on RA(1950): 23:53:22.9, Dec(1950): -68:36:46
(PKS catalogue, Gregory et al. 1994) and the optical spectrum
associated to these coordinates shows broad Ly-$\alpha$ emission at
4580 $\rm{\AA}$. This corresponds to a redshift $z\simeq 2.77$, see
Figure $\ref{fig:Lyalpha}$. Further investigation suggests that the
object at $z=1.716$ reported by Hewitt \& Burbidge is actually 115''
away at RA(1950): 23:53:28.3, Dec(1950): -68:35:24 (Hewitt \&Burbidge
1989).

\section{MgII absobers statistic} \label{mgII:sec}
\subsection{Redshift number density}

The number densities of strong
MgII absorption systems, $\partial N/ \partial z$, is simply defined as
\begin{equation}
\partial N/ \partial z=N/\Delta z,
\end{equation}
where $N$ is the number of absorbers in the redshift interval, $\Delta z$.
The errors on the number density is given by Poisson statistics, namely
\begin{equation}
\sigma_{\partial N/ \partial z} = \sqrt{N}/\Delta z. 
\end{equation}

\begin{figure}[tbh] 
\includegraphics[width=1.0\hsize]{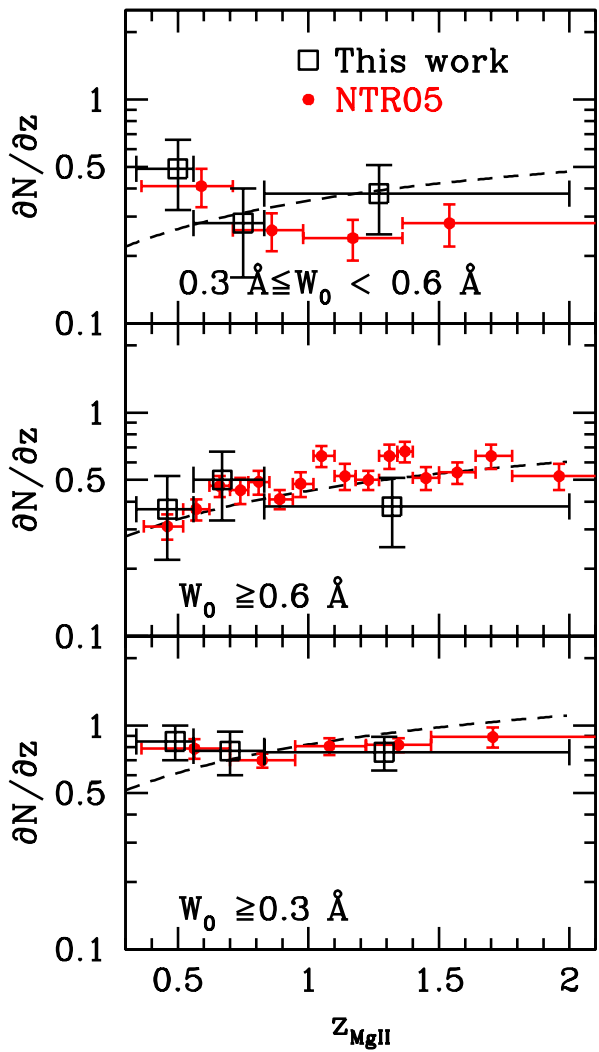}
\caption{Number densities of strong MgII absorption systems as a
  function of redshift obtained in this survey (black open squares)
  for different equivalent width ranges. Overplotted as dashed lines
  are the no-evolution curves, scaled with a $\chi^{2}$ fit to the
  observed redshift number densities. For comparison the number
  densities of Nestor et al. 2005 are plotted as red solid circles
  (NTR05).  Horizontal errorbars give the
  binsizes. \label{fig:dNvrsz_str}}
\end{figure}

Figure $\ref{fig:dNvrsz_str}$ shows $\partial N/ \partial z$ as
measured from our data (black open squares) as a function of redshift,
for three different values of the equivalent width, $W_{0}$. Our
results are found to be consistent with those of Nestor et al. (2005,
NTR05) from the SDSS survey (red filled circles) in the redshift range
$z=0.4-2.0$.  In particular, we find an almost constant $\partial
N/ \partial z \simeq 0.80\pm0.12$ at a mean redshift of
$z_{MgII}=0.85$, to be compared with $\partial N/ \partial z \simeq
0.783\pm0.033$ at a mean redshift of $z_{MgII}=1.11$, of NTR05 (see Table $\ref{tab:comp_strong}$).
Overplotted as dashed lines in Figure $\ref{fig:dNvrsz_str}$ are the
redshift number densites for a concordance $\Lambda$CDM universe with
parameters ($\Omega_{m}$,$\Omega_{\Lambda}$,$h$)=(0.27,0.73,0.7). In
these no-evolution curves (NEC) we assume a constant comoving number
density and a comoving cross sectional area proportional to
$(1+z)^{2}$. Note that in the equivalent width range $0.3 \le W_{0} < 0.6
\rm{\AA}$ the data also suggest an upturn of $\partial N/ \partial z$
compared with the NEC in the lowest redshift bin $z\approx 0.5$ which
NTR05 sees at $3\sigma$.

{\renewcommand{\arraystretch}{1.5}
\begin{table*}[tbh] 
    \centering
    \caption{Number densites of strong MgII absorption systems}
        \begin{tabular}{ccccc}%{l|l|l|l}
        \hline
        \hline 
        &\multicolumn{2}{c}{THIS WORK} & \multicolumn{2}{c}{NTR05}   \\
       %\hline
        
   $W_{0}^{2796}$ RANGE  & $\langle z_{MgII}\rangle$ & $\partial N/ \partial z$ & $\langle z_{MgII} \rangle$ & $\partial N/ \partial z$   \\
        \hline
        %$0.1 \le W_{0}^{2796}  < 0.3$ & $0.68$ &$0.82\pm0.12$  & & \\
         $0.3\mathring{\rm{A}} \le W_{0}^{2796} < 0.6 \mathring{\rm{A}}$ & $0.84$ &$0.38\pm0.08$ & &  \\
        $\ge0.3$ $\mathring{\rm{A}}$ & $0.85$ & $0.79\pm0.12$  &$1.11$ & $0.783\pm0.033$  \\
          $\ge0.6$ $\mathring{\rm{A}}$ & $0.84$& $0.41\pm0.09$ &$1.12$ & $0.489\pm0.015$  \\
          \hline
      \end{tabular}
    \label{tab:comp_strong}
    \begin{flushleft}
\end{flushleft}
\end{table*}}

\begin{figure*}
\includegraphics[width=0.80\textwidth]{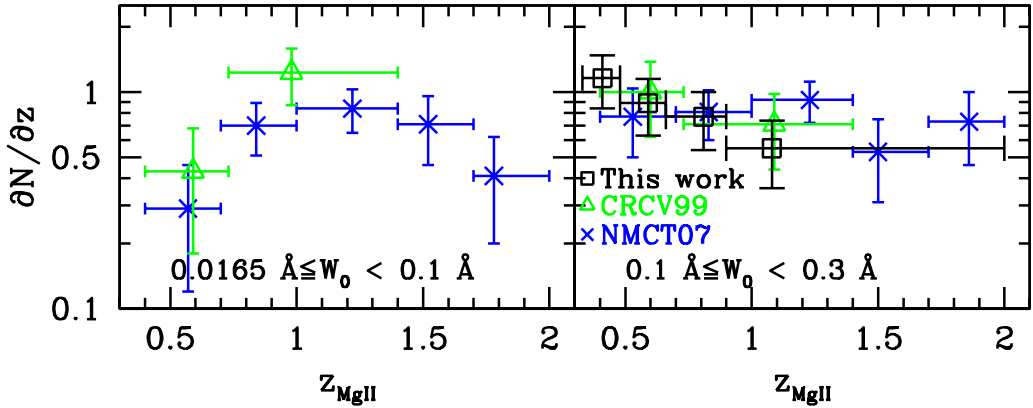}
\caption{Number densities of weak MgII absorption systems as a
  function of redshift for the equivalent width ranges 0.0165-0.1
  $\rm{\AA}$ and 0.1-0.3 $\rm{\AA}$. The values obtained in this work
  are shown by black open squares. For comparison the results of
  Churchill et al. 1999 (CRCV99) and Narayanan et al. 2007 (NMCT07)
  are overplotted as green empty triangles and blue crosses
  respectively. Horizontal errorbars give the
  binsizes. \label{fig:dNvrsz_w}}
\end{figure*}

Turning to the weak MgII absorption systems in Fig. $\ref{fig:dNvrsz_w}$,
we compare our results for $\partial N/ \partial z$ 
with those of Churchill et al. 1999 (CRCV99),
who looked for weak MgII absorption
lines in HIRES/Keck spectra of 26 QSOs in the redshift range 0.4-1.4,
and Narayanan et al. 2007 (NMCT07), who carried out a survey for weak MgII
absorbers using UVES/VLT archive data of 81 QSOs in the redshift range
0.4-2.4. 
In both works the $\partial N/ \partial z$ are given for
the equivalent width range $0.0165 \le W_{0} < 0.3 \rm{\AA}$.
Therefore
to compare with our result in the equivalent width range $0.1 - 0.3
\rm{\AA}$ we split their samples in 2 bins and recomputed the
$\partial N/ \partial z$ for the bins $0.0165 - 0.1 \rm{\AA}$ and $0.1
- 0.3 \rm{\AA}$. The evolution of $\partial N/ \partial z$ as a function
of redshift for these two bins is shown in Fig. $\ref{fig:dNvrsz_w}$ and the values of $\partial N/ \partial z$ are summarized in Table $\ref{tab:comp_weak}$. 

{\renewcommand{\arraystretch}{1.5}
\begin{table*}
    \centering
    \caption{Number densites of weak MgII absorption systems}
        \begin{tabular}{ccccccc}%{l|l|l|l}
        \hline
        \hline 
        &\multicolumn{2}{c}{THIS WORK} & \multicolumn{2}{c}{CRCV99} & \multicolumn{2}{c}{NMCT07}   \\
       %\hline
        
   $W_{0}^{2796}$ $\&$ z RANGE & $\langle z_{MgII}\rangle$ & $\partial N/ \partial z$ & $\langle z_{MgII} \rangle$ & $\partial N/ \partial z$ & $\langle z_{MgII} \rangle$ & $\partial N/ \partial z$    \\
        \hline
        $0.0165 \mathring{\rm{A}} \le W_{0}^{2796}  < 0.1 \mathring{\rm{A}}$, $0.4 \le z < 0.7$ &  &  & 0.59  & $0.43\pm0.25$ & 0.57  & $0.29\pm 0.17$ \\
        $0.0165 \mathring{\rm{A}} \le W_{0}^{2796}  < 0.1 \mathring{\rm{A}}$, $0.7 \le z < 1.4$ &  &  & 0.98  & $1.23\pm0.36$ & 1.07  & $0.78\pm 0.14$ \\
        $0.1 \mathring{\rm{A}} \le W_{0}^{2796}  < 0.3 \mathring{\rm{A}}$, $0.4 \le z < 0.7$ & $0.53$ &$0.94\pm0.21$  & 0.60  & $1.00\pm0.38$ & 0.53  & $0.77\pm 0.27$ \\
        $0.1 \mathring{\rm{A}} \le W_{0}^{2796}  < 0.3 \mathring{\rm{A}}$, $0.7 \le z < 1.4$ & $0.90$ &$0.65\pm0.16$  & 1.09  & $0.71\pm0.27$ & 1.06  & $0.87\pm 0.15$ \\
          \hline
      \end{tabular}
    \label{tab:comp_weak}
    \begin{flushleft}
    
  NOTES.- Redshift number densities $\partial N/ \partial z$ of weak MgII absorption systems for this work and the surveys of Churchill et al. 1999 (CRCV99) and Narayanan et al. 2007 (NMCT07). In these works $\partial N/ \partial z$ are given for the equivalent width range $0.0165 \le W_{0}^{2796}  < 0.3$ $\mathring{\rm{A}}$. Values of $\partial N/ \partial z$ given here for the two equivalent widths bins are computed values which assume the same redshift path for the two equivalent width ranges as for the whole range. This leads to small underestimates of $\partial N/ \partial z$ in the equivalent width range $0.0165 \le W_{0}^{2796}  < 0.1$ and small overestimates of $\partial N/ \partial z$ in the range $0.1 \le W_{0}^{2796}  < 0.3$ ( 2-4 \% in $\partial N/ \partial z$).\\
\end{flushleft}
\end{table*}}

Using the entire range of weak equivalent widths, both CRCV99 and
NMCT07 found a significant increase in $\partial N/ \partial z$ from
redshift 0.4 to 1.4.  However, after splitting in the two equivalent
width bins reported in Fig. $\ref{fig:dNvrsz_w}$, we find that this
behavior is dominated by systems in the equivalent width range $0.0165
- 0.1 \rm{\AA}$. In this range $\partial N/ \partial z$ increases from $0.43\pm0.25(0.29\pm0.17)$ to $1.23\pm0.36(0.78\pm0.14)$ from a mean redshift of $\langle z_{MgII}\rangle=0.59(0.57)$ to $\langle z_{MgII}\rangle=0.98(1.07)$ for the data of CRCV99(NMCT07). 

In the equivalent width range $0.1 - 0.3 \rm{\AA}$ and in the redshift range 0.4 - 1.4 we find that $\partial N/ \partial z$ decreases towards higher redshifts from $0.94\pm0.21$ at a mean redshift $\langle z_{MgII}\rangle=0.53$ to $0.65\pm0.16$ at $\langle z_{MgII}\rangle=0.90$. This finding is supported by our reanalysis of the CRCV99 data which gives a decrease of $\partial N/ \partial z$ of $1.00\pm0.38$ at $\langle z_{MgII}\rangle=0.60$ to $0.71\pm0.27$ at $\langle z_{MgII}\rangle=1.09$. The data of NMCT07 rather favor a flat $\partial N/ \partial z$ with $\partial N/ \partial z=0.77\pm0.27$ at $\langle z_{MgII}\rangle=0.53$ and $\partial N/ \partial z=0.87\pm0.15$ but the data are within the errorbars also still consistent with a decrease towards higher redshifts.

Assuming a power-law of the form $\partial N/ \partial z =
N_{0}(1+z)^{\gamma}$ for the redshift distribution of weak absorbers,
a formal $\chi^{2}$ fit leads to best fit parameters
$N_{0}=2.25_{-1.35}^{+2.85}$ and $\gamma=-1.9_{-1.6}^{+1.7}$. Thus our
observation suggest that for $z < 1.0$ we see an upturn in the
$\partial N/ \partial z$ for $0.1 \le W_{0} < 0.3 \rm{\AA}$ analog to
the one seen for the equivalent width range $0.3 \le W_{0} < 0.6
\rm{\AA}$.

\begin{figure*}
\includegraphics[width=0.80\textwidth]{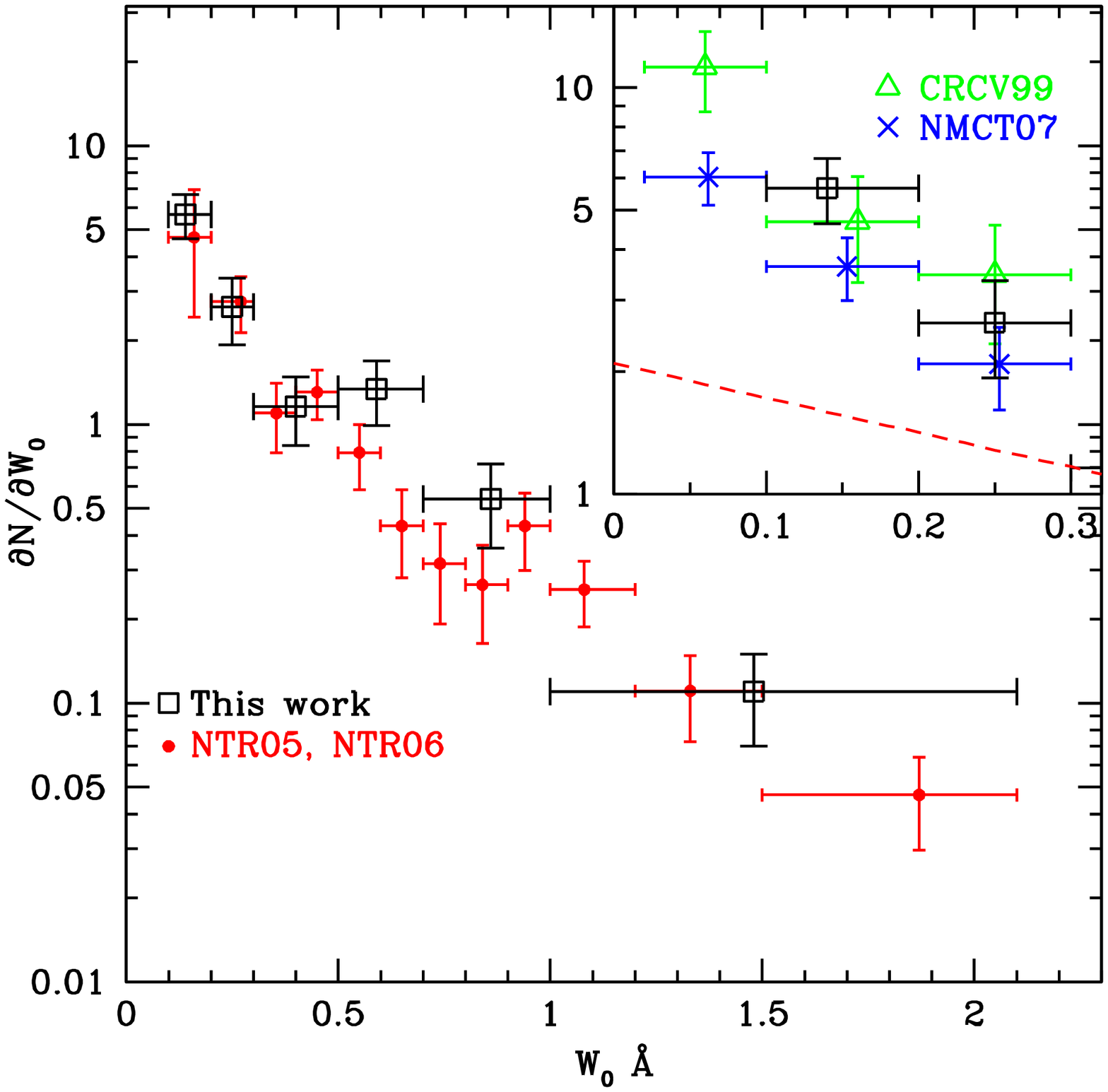}
\caption{Rest-frame equivalent width distribution function
  $\frac{\partial N}{\partial W_{0}}$, which is the number of MgII
  absorption systems with rest-frame equivalent width $W_{0}(2796)$
  per unit equivalent width per unit redshift, derived in this survey
  (black open squares). Overplotted with red solid circles is the
  equivalent width distribution function from Nestor et al. 2005
  ($W_{0} >= 0.3 \rm{\AA}$) and Nestor et al. 2006 ($W_{0} < 0.3
  \rm{\AA}$). For the weak absorbers also the results of Churchill et
  al. 1999 (CRCV99) and Narayanan et al. 2007 (NMCT07) are shown as
  green triangles and as blue crosses respectively. The red dashed
  line shows the fit of Nestor et al. 2005 of the form $\frac{\partial
    N}{\partial W_{0}}=\frac{N^{\ast}}{W^{\ast}}e^{-W_{0}/W^{\ast}}$
  with $N^{\ast}=1.187$ and $W^{\ast}=0.702$ to the systems with
  $W_{0} \ge 0.3 \rm{\AA}$.  \label{fig:distew}}
\end{figure*}

 \subsection{Equivalent width distribution}

 In Figure $\ref{fig:distew}$ the distribution of the rest-frame
 equivalent widths of the MgII systems is presented. Steidel $\&$
 Sargent 1992 (SS92) found that the number of MgII absorption systems
 with rest equivalent width $W_{0}(2796)$ per unit equivalent width
 per unit redshift can be either described by an exponential
 distribution of the form:
\begin{equation}
\vspace{+4pt}
\frac{\partial N}{\partial W_{0}}=\frac{N^{\ast}}{W^{\ast}}e^{-W_{0}/W^{\ast}},
\label{eq:expdist}
\end{equation}
with free parameters $N^{\ast}$ and $W^{\ast}$, or by a power law
distribution of the form:
\begin{equation}
\vspace{+4pt}
\frac{\partial N}{\partial W_{0}}=CW_{0}^{-\delta},
\label{eq:powlawdist}
\end{equation}

with free parameters $C$ and $\delta$. 

Using the Maximum-Likelihood method SS92 determined the parameters as
$N^{\ast}=1.55\pm0.20$, $W^{\ast}=0.66\pm0.11$ and $C=0.38\pm0.03$,
$\delta=1.65\pm0.09$ at a mean redshift of the absorbers of $\langle
z_{MgII}\rangle=1.12$. However, they noted that an exponential
distribution underpredicts the number of weak systems (here $ W_{0} <
0.5 \rm{\AA}$), whereas a power law fit overpredicts the number of
intermediate systems (here $0.7 \rm{\AA} < W_{0} < 1.3 \rm{\AA}$).

Our Maximum Likelihood fit of a single exponential function to MgII
systems with $W_{0} > 0.3 \rm{\AA}$ gives best fit parameters of
$W^{\ast}=0.74_{-0.14}^{+0.19} \rm{\AA}$ and
$N^{\ast}=1.18\pm0.33$. This is in very good agreement with the 
tightly constraint parameters of NTR05 with $W^{\ast}=0.702\pm0.017
\rm{\AA}$ and $N^{\ast}=1.187\pm0.052$ using over 1300 MgII doublets.

{\renewcommand{\arraystretch}{1.5}
\begin{table}[tbh] 
    \centering
    \caption{Best fit parameters $W^{\ast}$ and $N^{\ast}$ of the equivalent width distribution of the form $\frac{\partial N}{\partial W_{0}}=\frac{N^{\ast}}{W^{\ast}}e^{-W_{0}/W^{\ast}}$ for strong MgII systems.}
        \begin{tabular}{ccc}%{l|l|l|l}
        \hline
        \hline 
        
    & $W^{\ast} ( \rm{\AA}) $ & $N^{\ast}$   \\
        \hline
      THIS WORK    & $0.74_{-0.14}^{+0.19}$  & $1.18 \pm0.33$  \\
      NRT05   & $0.702\pm0.017$  & $1.187\pm0.052$ \\
      SS92 & $0.66 \pm0.11$  & $1.55\pm0.20$ \\
          \hline
      \end{tabular}
    \label{tab:comp_ewdist}
 %\begin{flushleft}
%\end{flushleft}
\end{table}}

For the equivalent width range $0.1\rm{\AA} - 0.3 \rm{\AA}$ we can
additionaly compare the $\partial N / \partial W_{0}$ with the results
of CRCV99 and NMCT07 which both give the $\partial N / \partial W_{0}$
in the redshift range 0.4 - 1.4. Below $0.3 \rm{\AA}$ the equivalent
width distribution distinctly rises above the single exponential fit
for $W_{0} \ge 0.3 \rm{\AA}$ as previously seen by Nestor et. al 2006
and NMCT07. This is further shown in the inset in Figure
$\ref{fig:distew}$ where the red dashed line corresponds to the best
fit to systems with $W_{0} > 0.3 \rm{\AA}$ by NTR05.  Note that there
is a slight discrepancy between NMCT07 and CRCV99, in that CRCV99
finds systematically more MgII systems in all 3 plotted equivalent
width bins ([0.0165,0.1], [0.1,0.2], [0.2,0.3] $\rm{\AA}$).  For the
lowest equivalent width bin, centered at $0.06 \rm{\AA}$, this is a
the level of $1.8 \sigma$. Unfortunately we are unable to make a
statement about the discrepancy because our redshift path falls dramatically in this equivalent width bin.

\section{Rotation Measure data}
 \label{RM:sec}
\subsection{Dependence of Rotation Measure distributions on equivalent width detection threshold}

\begin{figure*}
\includegraphics[width=0.80\textwidth]{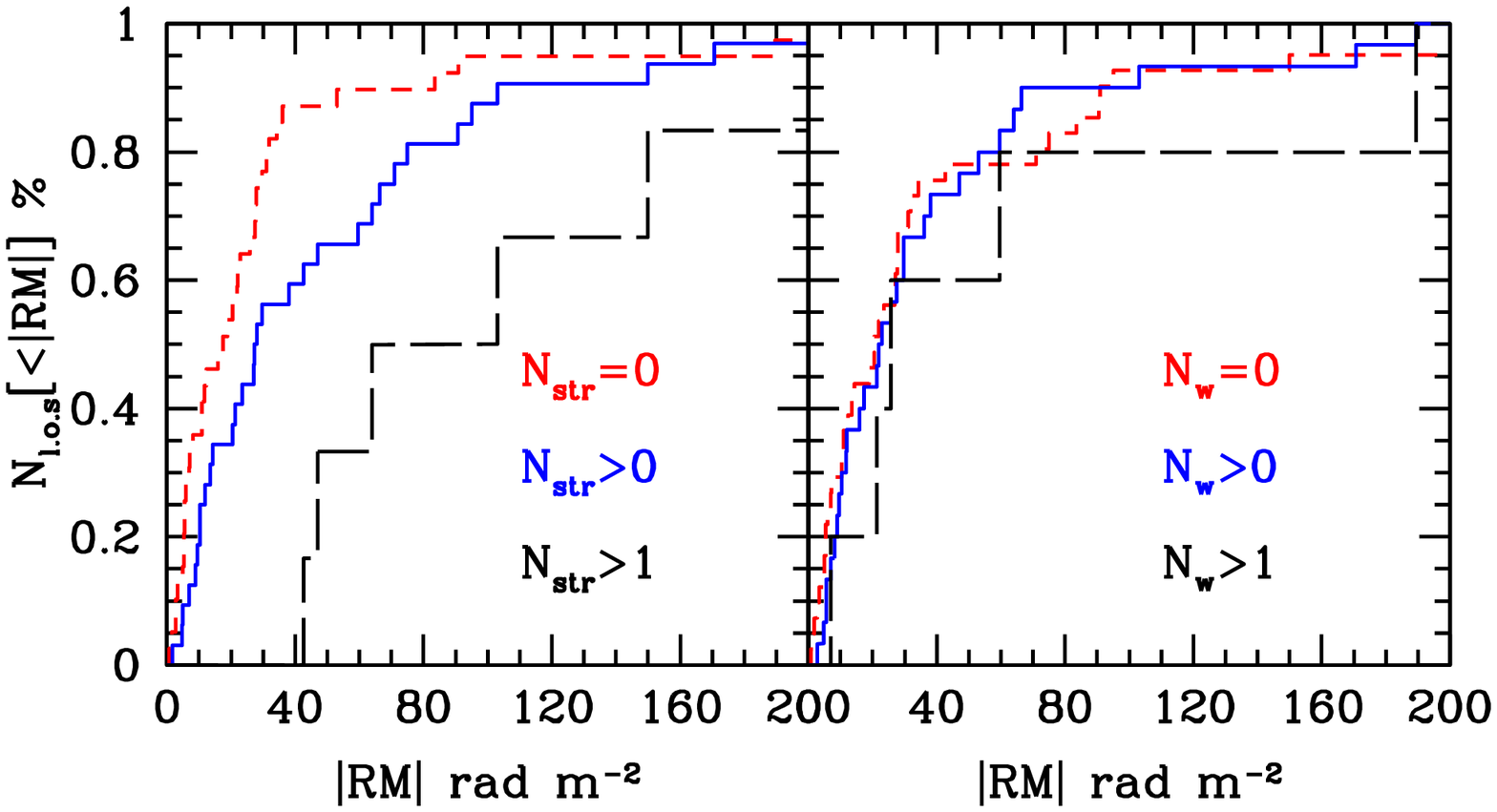}
\caption{Left panel: Cumulative Rotation Measure distribution
  functions for lines of sight having $N_{MgII}=0, > 0, >1$ strong
  MgII absorption lines (red short-dashed lines, blue solid lines,
  black long-dashed lines). Right panel: Same as left panel, but with
  the Rotation Measure distribution functions selected according to
  the number of weak MgII absorption
  lines.  \label{fig:Comp_strong_weak}}
\end{figure*}
\begin{figure*}
\includegraphics[width=0.80\textwidth]{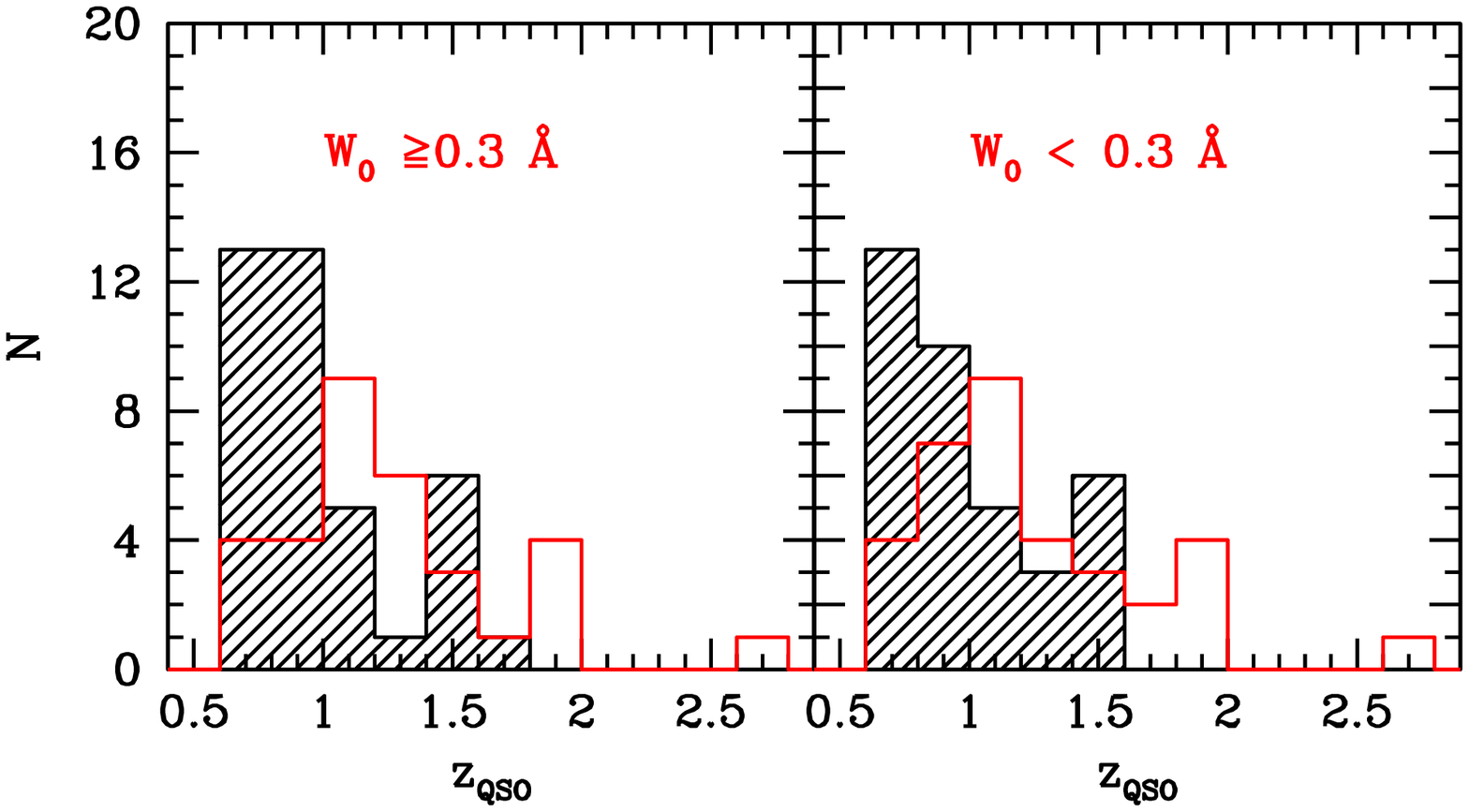}
\caption{QSO redshift distribution of l.o.s showing strong (left
  panel) and weak (right panel) MgII absorption (red histograms). The
  QSO redshift distributions of l.o.s which do not show MgII
  absorption in the corresponding equivalent width ranges are shown as
  black shaded histograms.  \label{fig:zQSO_weak_strong}}
\end{figure*}

Bernet et al. (2008)
showed that the Rotation Measure distribution function, for lines of
sight with strong MgII absorption lines, is significantly broader than
those without such lines. This led those authors to conclude 
that there must exist substantial magnetized plasma in or near the
absorption line systems. It is unclear at this stage whether the
association is direct, or whether the MgII absorption merely indicates
that the quasar sightline passes throug a galactic halo. Regardless,
this result suggests the presence of $\mu$Gauss-level large-scale
magnetic fields in or around typical galaxies when the Universe was a
half or less of its current age, with important implications for our
understanding of the development of magnetic fields in galaxies, and
in particular setting interesting constraints on the required
efficiency of the conventional galactic dynamo model.

Since our catalogue of MgII absorption systems is complete down to
$0.1 \rm{\AA}$, we can now repeat the analyis of Bernet et al. 2008
and test their result versus the applied equivalent width detection
threshold of MgII absorbers.  This would clarify whether or not weak
MgII systems also have large scale magnetic fields that contribute to
the observed Rotation Measure.

Figure, $\ref{fig:Comp_strong_weak}$, compares the RM
cumulative distribution functions having $N_{MgII}=0, > 0$, and $> 1$,
where $N_{MgII}$ refers to the number of strong and weak MgII
absorption lines for the left and right panel, respectively.  The left
panel is the result already presented in Bernet et al. (2008). For
this panel, a Kolmogorov-Smirnov test indicates that the RM distributions for
$N_{MgII}=0, N_{MgII} > 0$ are different at the 94.5$\%$ significance
level and those for $N_{MgII}=0, N_{MgII} > 1$ at the 99.98$\%$
significance level. Given that we were testing a specific hypothesis in an clean way with
completely independent observational data, we regard this result as significant. 
On the other hand, the right panel was built using the weak MgII
absorbers catalogue of this paper (including here also the 8 systems
below $W_{0} < 0.1 \rm{\AA}$) and the same RM data as in Bernet et
al. (2008). Note that in this case $N_{MgII}=N_w$, which does not
account for the number of strong MgII absorption lines along the lines
of sight. A KS-test does not recognize any difference in the RM distribution
functions for $N_{MgII}=0, > 0$ with a significance level of 22.1$\%$ and 31.0$\%$ for $N_{MgII}=0, > 1$. Since the RM distributions look statistically equivalent
independent of the number of weak MgII absorption systems, it is
immediately clear that the weak MgII absorption system contribution to
the observed QSO RM must be negligible. 

This result was also hinted at in observed redshift dependence of the
RM distribution (see Kronberg et al. 2008). In that work, our analysis
was based solely on the RM distribution of 268 quasars in the range $0
< z_{QSO} < 3.0$, without any spectroscopic information.  It was noted
that the observed increase in the width of the RM distribution with
redshift was better reproduced by a simple model in which the
statistics of the intervenors was given by the relatively rare strong
MgII systems than by one which included also the commoner weaker
absorption systems. Due to the increased number of intervenors at low
redshifts in this model, the increase in the width of the RM
distribution already starts at $z \approx 0.5$ which is not seen in
the actual data, where the increase happens around $z \approx 1$.

\subsection{Further evidence that strong MgII systems are responsible 
for the broadening of the RM distribution with z}

As pointed out in Bernet et. al (2008), because the probability of
intercepting a MgII absorber along the l.o.s increases strongly with
the QSO redshift, a strong evolution of the magnetic fields, 
local to the QSOs could give rise to a fitticious
correlation of $\vert \rm{RM} \vert$ with the number of strong MgII
absorption lines.  The evolution in the magnetic fields local to the
QSOs would have to be significant, in order to produce the observed
increase in RM dispersion with z despite the strong
$\propto (1+z)^{-2}$ ``k-correction'' in the observed RM, 
due to the $\lambda^{-2}$ dependence of the RM.

By comparing the median of the $\vert \rm{RM} \vert$ distribution of
l.o.s with and without strong MgII absorption for different QSO
redshifts, Bernet et al. (2008) found that there is only a 7$\%$
chance probability that their result was due the above spurious
correlation or, in general, to a correlation between redshift and
$\vert \rm{RM} \vert$.

Our new demonstration above that sightlines with weak MgII absorption
do not show statistically enhanced $\vert \rm{RM} \vert$ values,
can be used as an even more convincing test
to rule out the possibility that an underlying correlation between
$\vert \rm{RM} \vert$ and $z_{QSO}$ causes the observed RM broadening
at high redshifts.

\begin{figure*}
\includegraphics[width=0.80\textwidth]{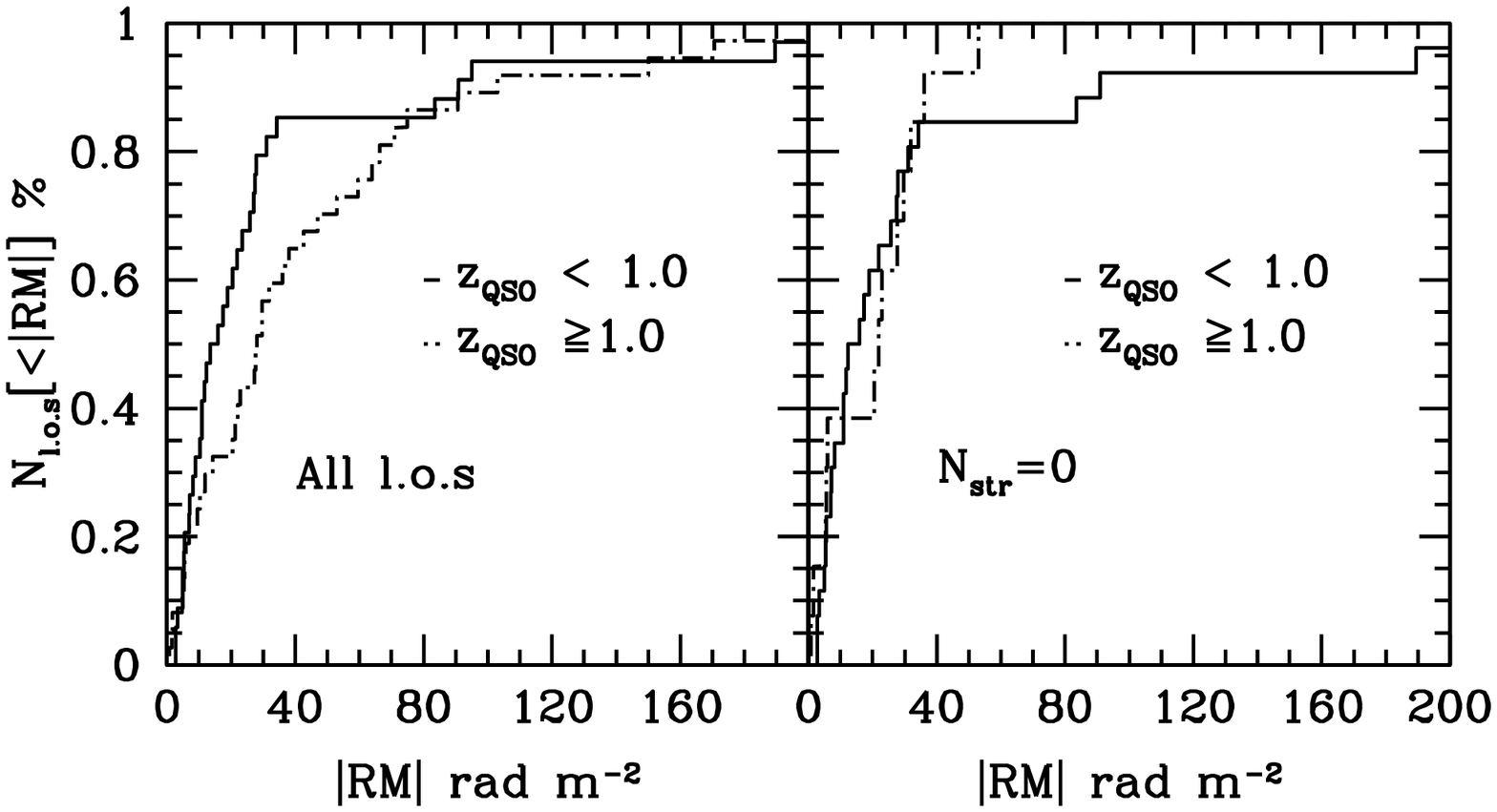}
\caption{Left panel: Cumulative Rotation Measure distribution
  functions for lines of sight with QSO redshift $z_{QSO} < 1.0$
  (solid line) and $z_{QSO} \ge 1.0$ (dashed-dotted line). Right
  panel: Same as left panel but only for lines of sight which do not
  show strong MgII absorption systems.  \label{fig:DistRM_low_highz}}
\end{figure*}

In the left-hand panel of Figure
$\ref{fig:zQSO_weak_strong}$ the redshift distributions for the
QSOs whose sightlines do and do not exhibit strong MgII absorption
systems are shown as red and black shaded histograms. The same is
shown on the right-hand panel for the weak MgII absorption
systems. Apparently the QSO redshift distributions are
very similar independent of whether the MgII
absorption systems belongs to the weak or strong category.
A KS-test reveals no difference between them with a significance level of $1.21\%$ 
for sightlines with weak and strong MgII absorption systems and $0.01\%$ for sightlines 
without weak or strong MgII absorption systems.
 
The cumulative $\vert \rm{RM} \vert$ distributions for different
number of MgII systems in Figure $\ref{fig:Comp_strong_weak}$ look,
however, very different, depending on whether they are selected
according to the presence of strong or weak MgII systems. There is a
clear broadening of the $\vert \rm{RM} \vert$ distribution with
increasing number of strong MgII absorption lines $N_{str}$. However,
there is virtually no difference in the $\vert \rm{RM} \vert$
distributions for different numbers of weak absorbers $N_{w}$. This
rules out the possibility that the correlation between $\vert \rm{RM}
\vert$ and the number of strong MgII absorption lines observed in
Bernet et. al (2008) is due to an underlying correlation between
$\vert \rm{RM} \vert$ and $z_{QSO}$.

Figure $\ref{fig:DistRM_low_highz}$ shows further
evidence that the broadening in the RM distribution with redshift is
caused by the magnetic fields traced by strong MgII systems.
The left-hand panel compares
the cumulative RM distribution function for QSO from our entire sample
with $z \ge 1.0$ and $z < 1.0$.  It clearly shows that the high
redshift distribution function is broader with respect to the
low redshift counterpart. This is most likely due to the fact that 
the chances of intercepting a MgII absorption systems are higher
at higher redshifts.
In fact, as we take out the lines of sight
containing strong MgII absorbers, the two distributions do not show
significant differences anymore (right panel). For the lines of sight
without strong MgII absorbers, the median $\vert \rm{RM} \vert$ for
both low ($\bar{z}_{QSO}=0.79$) and high redshifts
($\bar{z}_{QSO}=1.42$) is around 20 rad $\rm{m^{-2}}$. This suggests
that the l.o.s without strong MgII absorption systems (39 QSOs) are
dominated by contributions from magnetic fields within the Milky Way,
which is about 20 rad m$^{-2}$ (Bernet et al. 2008).

Any
extragalactic RM contributions which increase with z,
e.g. intergalactic magnetic fields, envolving RM contributions local
to the QSO, are probably swamped by the Galactic one. It is
interesting to note that all 4 $\vert \rm{RM} \vert$ values with
$\vert \rm{RM} \vert > 100$ rad $ \rm{m^{-2}}$ are at $z < 1.0$. This
might be partially explained by the fact that a RM contribution local
to a QSO at redshift $z_{QSO}$ is reduced by a factor
$(1+z_{QSO})^{-2}$ when transformed to the observers frame. Thus
(non-evolving) RM contributions local to the QSO might just fall below
the Galactic contribution at significantly higher redshifts than 1.

\subsection{Why do strong MgII systems contribute to the observed Rotation Measure and weak ones not?} 

In order to address the question of why weak absorbers do not
contribute to the RM of distant QSOs, it is important to understand
the nature of the weak MgII absorber systems. It has been proposed that these systems are associated with low surface brightness galaxies (Churchill $\&$ Le Brun 1997), intergalactic star forming pockets (Rigby et al. 2002) or dwarf galaxies (Zonak et al. 2004). More recent works (Churchill et al. 2005, Kacprzak et al. 2007), however, indicate that for some fraction of the weak MgII systems a normal $(L=0.1 -10 L_{B}^{\star})$ associated galaxy can be found. The impact parameters of the 7 weak MgII system where Churchill et al. 2005 have HST imaging range from 35 to 115 kpc. It still needs to be sorted out how this is consistent with the general picture that weak MgII systems are sub Lyman Limit systems.

The lack of quantitative knowledge about the fraction of weak MgII systems that can be associated with normal galaxies like the ones traced by the strong systems makes it hard to draw strong conclusions from our observation. We need to have a better knowledge about the differences in the impact parameter distribution of the galaxies traced by the strong and weak systems and their properties, e.g. luminosity, color. The work of Chen $\&$ Tinker 2008 shows that the impact parameters of the galaxies traced by the weak absorbers are generally larger than those traced by the strong systems; they find a moderate anticorrelation between the equivalent width of the identified MgII systems and the impact parameters of the galaxies at the 97$\%$ level. An inspection of the impact parameters in Chen $\&$ Tinker 2008 shows that the median impact parameters of the weak absorbers is around $D \approx 60 \rm{kpc} $ whereas $D \approx 40 \rm{kpc} $ for the strong absorbers.

It is thus plausible that the weak systems are preferentially
produced by l.o.s through the outer regions of a galaxy. Since RM
$\propto n_{e}B$, with $n_{e}$ the density of free electrons and B the
magnetic field along l.o.s, we also expect any RM contribution to
decline quite rapidly with impact parameter.
For strong absorption systems, it has been suggested that on average
larger equivalent widths (i.e. stronger absorbers) correspond to
galaxies with bluer spectra and smaller impact parameters (Zibetti et
al 2007).  If this trend was found to also extend to the weak
absorbers, one could also attribute the presence of weaker magnetic
fields in weak absorbers to the lower star formation activity there.

\section{Summary} \label{Sum:sec}

We have presented a catalogue of strong and weak MgII absorption
systems with equivalent width down to 0.1 $\rm{\AA}$, obtained from a
survey of 77 QSOs using the UVES spectrograph at the VLT.  We
determine the statistical properties of strong MgII systems and find
them in good agreement with previous results. In particular, we
confirm the upturn in $\partial N/ \partial z$ at lower redshifts (z
$<$ 0.6) for systems in the equivalent width range $0.3 \le W_{0} <
0.6 \rm{\AA}$. Compared with no-evolution models in a $\Lambda$CDM
universe we find more MgII systems at lower redshifts.  

Concerning the weak absorption systems, we point out that the
previously observed increase with redshift of $\partial N/\partial z$
(CRCV99, NMCT07), pertains only to the very weak absorbers with $W_{0}
<0.1 \rm{\AA}$. Instead, $\partial N / \partial z$ for absorbers with
$W_{0}$ in the range 0.1--0.3 $\rm{\AA}$ actually decreases, similarly
to the case of strong absorbers.

We use this catalogue to extend our previous analysis of the
connection between the presence of intervening absorption and the
Faraday Rotation Measure of the quasar.
We show that unlike strong MgII systems, weak MgII absorbers do not
contribute to the observed Rotation Measure.  This is likely due to
the higher impact parameters of, and/or to lower star formation
activity systems traced by, the weak absorbers with respect to 
strong ones.  We use the lack of correlation of RM with number of weak
MgII absorbers to rule out the possibility that the correlation of RM
with number of strong MgII absorbers observed in Bernet et al. (2008)
is due an underlying correlation of Rotation Measure with redshift,
caused, for example, by strong evolution of magnetic field local to
the QSOs environments.

We also show that, while the distribution of RM for QSOs above $z=1$
is distinctly broader than the corresponding distribution for QSOs
with $z<1$, the difference disappears once the lines of sight
exhibiting MgII absorption are removed.  This further shows that: (a)
the increase in the width of the RM distribution with redshift is
indeed caused by large scale magnetic fields traced by strong MgII
systems and (b) that any further extragalactic RM contribution is most
likely swamped by local Milky Way foreground contributions.
 
\section{Acknowledgements}
M.L.B. acknowledges financial support from the Swiss National Science
Foundation.
This research has made use of the NASA/IPAC Extragalactic Database
(NED) which is operated by the Jet Propulsion Laboratory, California
Institute of Technology, under contract with the National Aeronautics
and Space Administration

\end{document}